\documentclass{ws-procs9x6}
\usepackage{epsfig}
\newcommand{\be}{\begin{equation}}
\newcommand{\ee}{\end{equation}}

\newcommand{\vp}{\varphi}
\begin{document}

\title{Quantum field theory approach to the vacuum replica in QCD}

\author{A.~V. NEFEDIEV\footnote{\uppercase{W}ork
supported by \uppercase{INTAS} via grants \uppercase{OPEN}
2000-110 and \uppercase{YSF} 2002-49 and by \uppercase{RFFI} via
grants 00-02-17836, 00-15-96786, and 01-02-06273.}}

\address{Centro de F\'\i sica das Interac\c c\~oes Fundamentais,\\
Departamento de F\'\i sica, Instituto Superior T\'ecnico, \\
Av. Rovisco Pais, P-1049-001 Lisboa, Portugal\\
and\\
Institute of Theoretical and Experimental Physics, \\
117218, B.Cheremushkinskaya 25, Moscow, Russia}

\author{J.~E.~F.~T. RIBEIRO}

\address{Centro de F\'\i sica das Interac\c c\~oes Fundamentais,\\
Departamento de F\'\i sica, Instituto Superior T\'ecnico, \\
Av. Rovisco Pais, P-1049-001 Lisboa, Portugal}

\maketitle

\abstracts{ Quantum field theory is used to describe the
contribution of possible new QCD vacuum replica to hadronic
processes. This sigma-like new state has been recently shown to be
likely to appear for any realistic four-quark interaction kernel
as a consequence of chiral symmetry. The local operator creating
the replica vacuum state is constructed explicitly. Applications
to physical processes are outlined.}

It was found recently that the fermionic vacuum of QCD is very
likely to possess a replica\cite{BNR}. The first evidence for such a state 
was given long ago when the oscillator-type confining force was found to
support and infinite tower of excited vacuum states\cite{Orsay2}.
A further analysis demonstrates that the very existence of
the replica depends both on the dimensionality of the
space-time and on the strength and the form of the interaction,
and, for a class of  realistic chiral models described below, one, and only
one, replica was shown to exist. Furthermore, the existence of such a replica
was numerically shown to be  stable against variations of the model 
parameters\cite{BNR}. The theory is given by the Hamiltonian 
\be 
H=\int
d^3 x\bar\psi(-i\vec{\gamma}\vec{\bigtriangledown})\psi+ \int
d^3xd^3y \left[\bar\psi\gamma_\mu\frac{\lambda^a}{2}\psi\right]_x
K_{\mu\nu}^{ab}(\vec{x}-\vec{y})
\left[\bar\psi\gamma_\nu\frac{\lambda^b}{2}\psi\right]_y 
\label{H}
\ee 
with the current-current interaction defined by the kernel
$K_{\mu\nu}^{ab}$, for which various forms can be considered
ranging from delta-functional\cite{NJL} and
oscillator-type\cite{model,BR} to more realistic linear form (see,
for example, \cite{linear}). Following the paper\cite{BNR}, we
consider the kernel which incorporates the linear confinement, the
colour-Coulomb interaction, and the constant term: 
\be
K_{\mu\nu}^{ab}(\vec{x}-\vec{y})=\delta^{ab}g_{\mu\nu}\left[g_{\mu0}\sigma_0|\vec{x}-\vec{y}|
-\frac{\alpha_s}{|\vec{x}-\vec{y}|}\left(1-e^{-\Lambda|\vec{x}-\vec{y}|}\right)+U\right].
\label{kernel} 
\ee 
It can be argued that such an interaction is
fixed by the only dimensional parameter given by the
nonperturbative scale, $\Lambda\sim U\sim\sqrt{\sigma_0}$, which
takes the standard value, and so does the strong coupling
constant: $\sqrt{\sigma_0}\sim 300MeV$, $\alpha_s=0.3$\cite{BNR}. 
The standard Bogoliubov technique is employed to
define the dressed quarks parameterized by the chiral angle
$\vp(p)$. The vacuum energy is to be minimal for the true vacuum
state, and this yields the mass-gap equation $\delta E_{\rm
vac}/\delta\vp=0$. In Fig. 1 we give the profiles of the solutions
to the mass-gap equation for the kernel (\ref{kernel}). Both
nontrivial solutions, $\vp_0$ and $\vp_1$, were shown to be extremely
resistant against variations of the interaction, and no more
solutions were found. The mere existence of such a replica, as described by 
the second solution $\vp_1$, is enough for what follows without the need 
to refer to any given particular model .

\begin{figure}[t]
\centerline{\epsfxsize=5.5cm\epsfbox{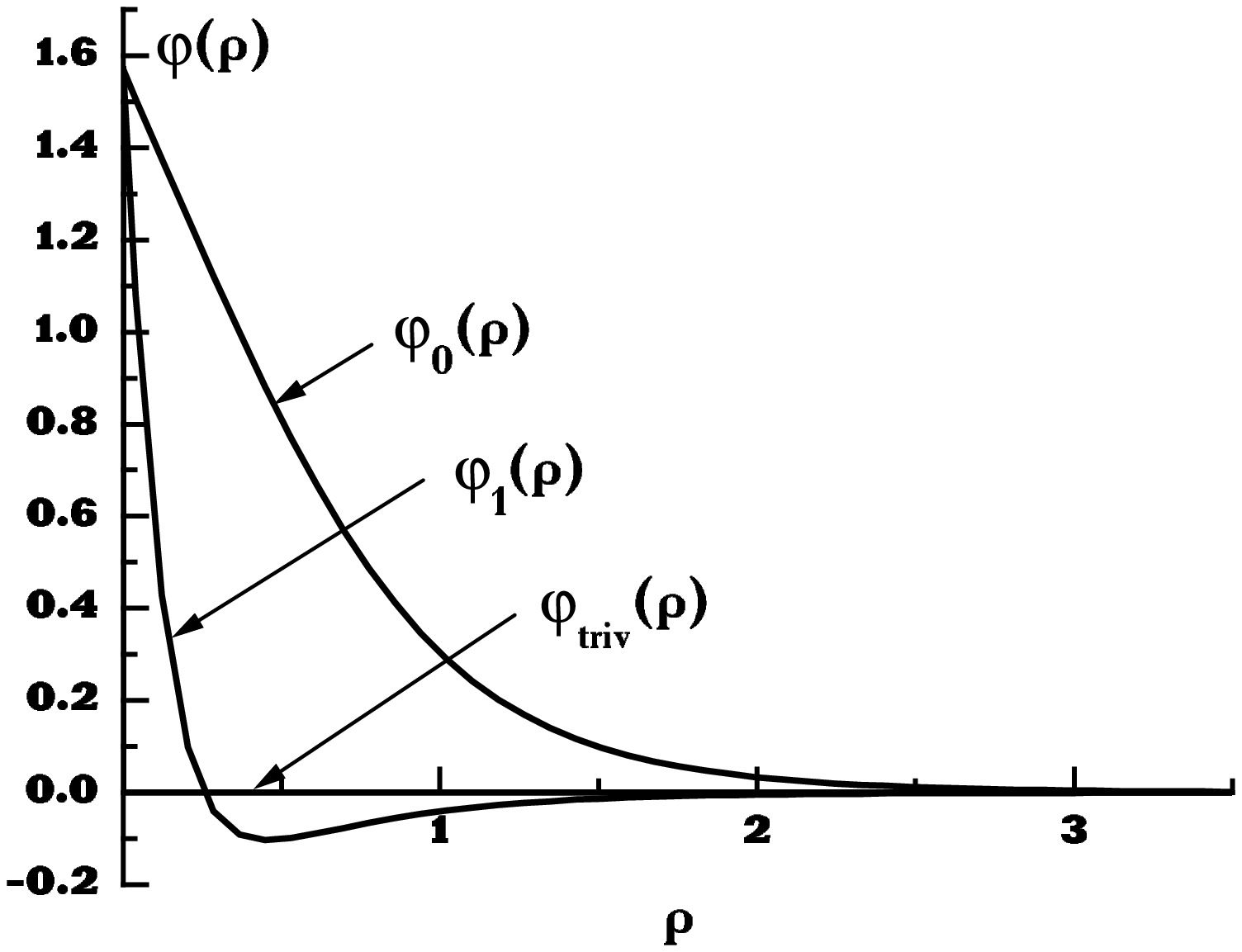}\epsfxsize=5.5cm\epsfbox{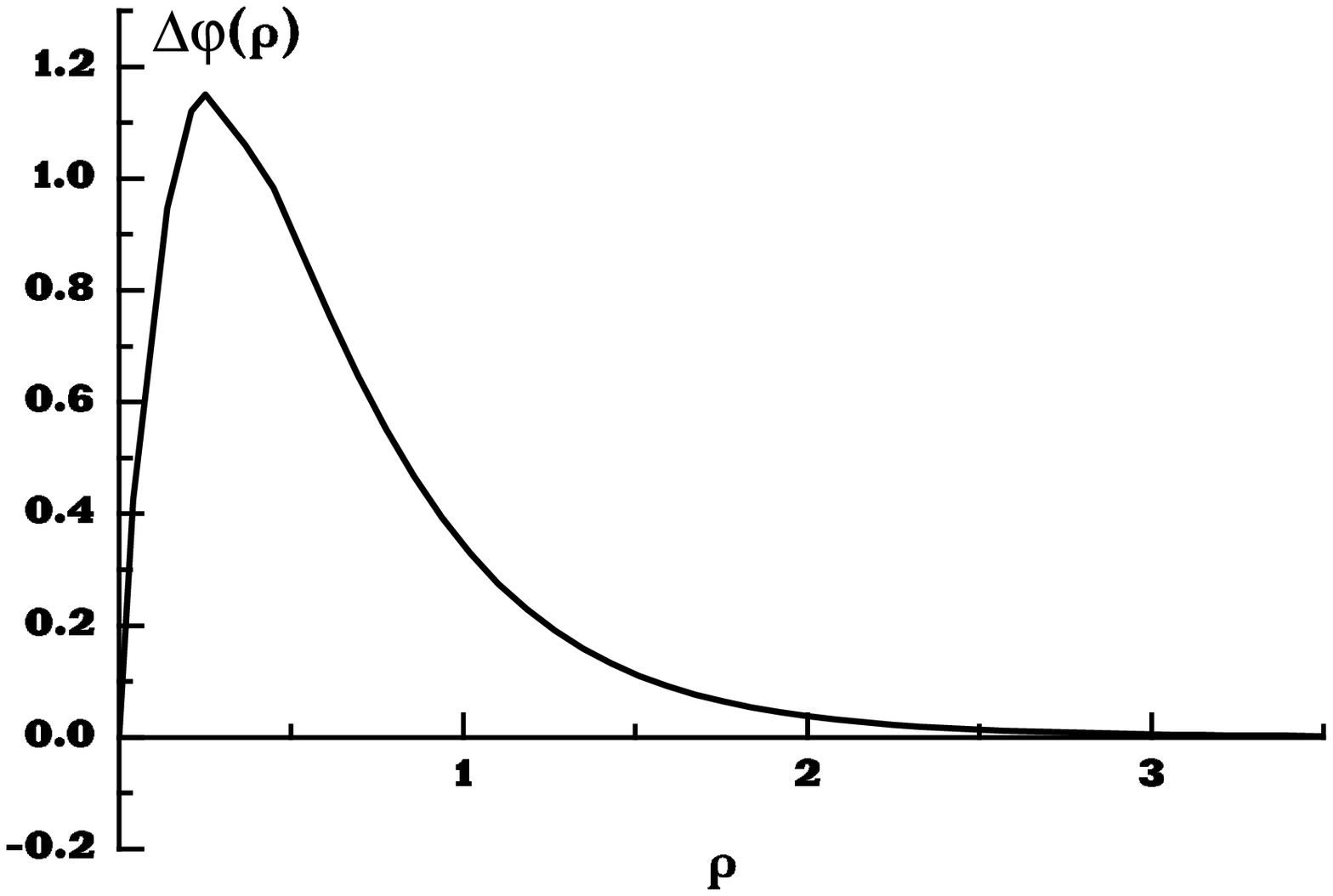}}
\vspace*{-5mm}
\caption{Three solutions to the mass-gap equation
and the difference $\Delta\vp=\vp_0-\vp_1$\protect\cite{BNR}.}
\end{figure}

The operator connecting the BCS vacuum with the replica
is\cite{BR}
\be
|1\rangle=e^{\Delta Q^\dagger-\Delta
Q}|0\rangle\equiv S|0\rangle, \quad \Delta
Q^\dagger=\frac12\int\frac{d^3p}{(2\pi)^3}\Delta\vp
C_{pcf}^\dagger,
\label{dq}
\ee
where the operator
$C_{pcf}^\dagger$ creates a pair of the quark and the antiquark of
a given colour and flavour with the vacuum quantum numbers $0^+$
and with the relative momentum $2\vec{p}$. Using Eq.~(\ref{dq})
one can show that the two vacua are orthogonal. For the
unitary operator $S$ from (\ref{dq}) we find the following
representation in terms of the quark fields:
\be
S=T\exp\left[\int
d^4x
f(x_0)\bar\psi(\vec{x},x_0)(-i\vec{\alpha}\vec{\bigtriangledown})
\psi(\vec{x},x_0)\right],
\label{Sop}
\ee
where the Fourier
transform of the even function $f(x_0)$ is
$f(\omega)=f(-\omega)=\frac{\Delta\vp(p_*)}{2p_*}$, $p_*$ being
such that $E_{\rm quark}(p_*)=\omega/2$. The operator (\ref{Sop})
creates a local excitation of the vacuum, $f$ playing the role of
an external field. Thus we define the following Feynman rules:
each f-vertex is to be supplied by
$f(\omega)(\vec{\alpha}\vec{p})$ with $\omega$ and $\vec{p}$ being
the energy pumped to the system in this vertex and the floating by
three-dimensional momentum, respectively. There is an integral
over the energy $\omega$. The full diagram is to be supplied by an
overall energy conservation $\delta$-function and multiplied by
the time density of the replica excitations, $n=N/T$, which
accounts for the fact that during the full time of the hadronic
process $T$ the replica has been excited $N$ times. These rules
can be used to consider corrections to hadronic processes due to
the replica excitations at the intermediate stages of the
reaction. For example, as the first check of selfconsistency, 
one can rederive the orthogonality condition for the two vauum states:
\begin{eqnarray}
\langle 0|1\rangle=1-V\int\frac{d^4p}{(2\pi)^4}\frac{d\omega}{2\pi}
f(\omega)f(-\omega)
\left[(\vec{\alpha}\vec{p})S(p_0,\vec{p})(\vec{\alpha}\vec{p})
S(p_0+\omega,\vec{p})\right]+\ldots\nonumber\\
=1-V\int\frac{d^3p}{(2\pi)^3}\left(\frac{\Delta\vp}{2}\right)^2+\ldots
=\exp\left[V\int\frac{d^3p}{(2\pi)^3}\ln\left(\cos^2\frac{\Delta\vp}{2}\right)
\right]\mathop{\to}\limits_{V\to\infty}0.\nonumber
\end{eqnarray}

In other words, we predict the existence of a new scalar
particle-like object with the dominating hadronic mode of decay
being the two-pion mode and which exists in the theory {\it
together} with the standard $\sigma$-meson --- the solution of the
bound-state Bethe-Salpeter equation\cite{NR}.


\begin{thebibliography}{0}
\bibitem{BNR} P. Bicudo, A. Nefediev, J. E. Ribeiro, {\it Phys.
Rev.} {\bf D65}, 085026 (2002).
\bibitem{Orsay2} A. Le Yaouanc, {\it et al.}, {\it Phys. Lett.}
{\bf B134}, 249 (1984).
\bibitem{NJL} Y. Nambu, G. Jona-Lasinio, {\it Phys. Rev.} {\bf 122}, 345
(1961).
\bibitem{model} A. Le Yaouanc, {\it et al.}, {\it Phys. Rev.} {\bf D29}, 1233 (1984);
{\it ibid.}, {\bf D31}, 137 (1985).
\bibitem{BR} P. Bicudo, J. E. Ribeiro, {\it Phys. Rev.} {\bf D42}, 1611, 1625, 1635
(1990).
\bibitem{linear} S. L. Adler, A. C. Davis, {\it Nucl. Phys.} {\bf B244}, 469 (1984);
N. Brambilla, A. Vairo, Phys. Lett. B {\bf 407}, 167 (1997); Yu.
A. Simonov, {\it Yad. Fiz.} {\bf 60}, 2252 (1997). F. J.
Llanes-Estrada, S. R. Cotanch, {\it Phys. Rev. Lett.}  {\bf 84},
1102 (2000).
\bibitem{NR} A. Nefediev, J. E. Ribeiro, in preparation.
\end{thebibliography}
\end{document}